# Confocal micro-Raman Spectra of untreated and lethally treated *Escherichia coli* exposed to UV-B and violet light.

## WERDEN KEELER[1,3] AND KAM LEUNG[2,4]

[1]*Physics Department, Lakehead University, 955 Oliver Road, Thunder Bay, Ontario, Canada P7B 5E1*

[2]*Biology Department, Lakehead University 955 Oliver Road, Thunder Bay, Ontario, Canada P7B 5E1*

[3]*wjkeeler@lakeheadu.ca*

[4]*kleung@lakeheadu.ca*

**Abstract:** We report on changes in the Raman spectrum of live *Escherichi coli (E.coli)* that result from exposure to lethal fluences of 300 nm (UV-B) and 405 nm (violet) photons. In the first instance, the energy per photon of 4.13 eV is sufficient to break most of the inter-atomic bonds in the bacterium and major change in the Raman spectrum, particularly in the RNA/DNA regions is observed. This energy is in near resonance with the C-H, N-H, and P-O bond binding energies that interconnect phosphate backbone sections and may initiate the Raman modifications. By contrast, the 3.06 eV violet photon energy is insufficient to cleave the stronger system bonds. The much larger lethal fluence required in this case produces a significant change in the resonant C-N and C-P bond connected amino acid/protein groups and lipid peak signal but much less so in the nucleotide-DNA/RNA signature.

## 1. Introduction

Ultraviolet light has been used extensively as a bactericidal agent for many years [1-4] and even mid-visible light can induce cellular mortality if the dose is high enough [5,6]. In an earlier article the requisite integrated fluence of light for mortality of *E.coli* using wavelengths from 250 nm to 488 nm was quantified [7]. From results of that work it was shown (see Fig.1a) that the required mortal fluence of photons grew close to exponentially with increase in wavelength ranging from 16 mJ/cm$^2$ at 275 nm wavelength to ~7.6x10$^7$ mJ/cm$^2$ at 488 nm. In all cases of treatment in that study, the optical power density was kept low enough that no heating of the sample was observable. An equation predicting the surviving fraction after a particular exposure was also derived and can be used to estimate the cumulative treatment effect for either a series of monochromatic input beams or for a spectral distribution of photons as from a Planck source for example. While the integrated exposure required at longer wavelengths was much greater, these lower energy photons, now readily obtainable from inexpensive semiconductor diode lasers for example, were still capable of complete cellular mortality. While visible wavelength light might appear at first glance to be ineffectual when dealing with organic processes, hydrogen bonding, protein folding mechanisms, and thermally mediated processes in living cells occur at approximately 100 fold lower energies so one should perhaps be less surprised by the results.

While the bactericidal fluences are now quite well known, there is still a dearth of spectroscopic information detailing the biological effects induced at the molecular level by these radiations. The importance of having light as a reliable killing agent in the microbial world with antibiotic resistant bacterial strains on the increase and viruses a growing problem, underpins the need for a better understanding of the killing mechanisms involved as a function of wavelength. In particular, killing with lower energy visible photons is of interest because of the ready availability of such light sources and their higher penetration ability. An examination of the equivalent photon energy for the dominant bond types found in organic molecules [19] is presented in Fig. 1(b). A given wavelength may materially affect bond integrity for all bond energies lower than that of the incident photon but may have the strongest impact on bond energies resonant or nearly resonant with the radiation energy.

Determining the changes that occur in the biological makeup of the bacterial cells as the treatment wavelength is lengthened and the required mortal fluence increases by so many orders of magnitude, is of major importance. A photon of 300 nm wavelength can in principle cleave most of the single bonds in a living cell, (O-H being the exception) through photo-electron ejection for example. XPS studies of damage induced by UV radiation on DNA [8] report release of CO and nitrogen based compounds through bond breakage, for energies above 4.2 eV. As the treatment wavelength lengthens however, fewer and fewer bond groups <span>B, **2015**, *119* (17), pp 5404–5411</span>are at high risk of damage by individual photons. Broadly speaking however, one might expect that the relatively weak C-N bond (3.12 eV) that binds amino acids together and ties nucleic acids to the phosphate backbone, to still be vulnerable. This bond energy is represented by a violet photon of about 400 nm wavelength. Because nucleic acid integrity and protein stability are both essential conditions for living organisms, we chose conveniently available 300 nm and 405 nm light sources for treatment investigations in this study.

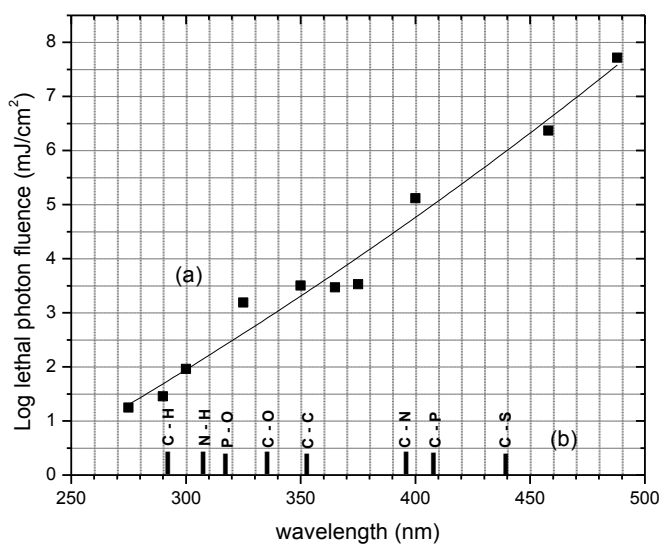

**Figure 1**. (a) Semi-log plot of the lethal fluence of light in millijoules per cm² of exposed cell suspension at a cell density of 7.8x10⁷ CFU/ml as taken from [7]. The line is a best least squares fit to the data. The uncertainty per point is equivalent to the size of the square in each case. (b) Bond energy of various organic bond types expressed in equivalent photon wavelengths.

To investigate changes in the structure of the *E.coli* cells, Raman spectroscopy was chosen because of its low sensitivity to the presence of water, sharp spectral features, well documented Raman spectra database for biological molecules [9,10] and ease of sample preparation. The method has previously been used to identify single bacteria cells [11,12], determine their chemical composition [13,14] and compare different strains of bacteria [15,16]. Raman spectroscopy has been applied to monitor the action of antibiotics to resistant and non-resistant bacterial strains [17,18] but to our knowledge, has yet to be used to monitor changes in bacterial molecular vibration spectra and hence structure, due to treatment with light or other electromagnetic radiation.

## 2. Materials and Methods

### 2.1 Bacteria Preparation

The JM109 *E.coli* strain, a Gram-negative bacterium, was obtained from the American Type Culture Collection (ATCC). This bacterium is a non-pathogenic strain and a common lab strain used for molecular cloning. The bacterial culture was stored in 25% glycerol at -80C and re-grown on Trypticase Soy Agar (TSA) (Difco, Detroit, MI, USA) plates at 30C *E.coli* cells taken from the TSA plates were then cultured in Trypticase Soy broth (TSB) (Difco) in a shaking incubator at 200 rpm at 30C. After 12-15 hours of incubation, the culture was centrifuged for 2 minutes at 3000xg, and then washed and centrifuged in 0.85% sterile saline solution 3 more times. The washed cells were suspended in the same solution and set at an OD concentration reading of 0.3 (7.85 x $10^7$ CFU/mL) using an optical densitometer, prior to UV and visible radiation treatments. This liquid concentration provided adequate cell numbers for accurate plating statistics and was still transparent enough to ensure that all cells received equal exposure fluence during treatment in a shallow container.

### 2.2 Confocal Micro-Raman Spectrometer

Micro-Raman spectra of *E.coli* both live and bactericidally treated with light, were obtained using a proprietary instrument assembled in house. In brief, the system consisted of a Nikon Eclipse E400 upright fluorescence microscope with the illuminator replaced by external coupling optics for an $Ar^+$ ion laser source. The laser input beam was centered and focused so that the laser spot and sample were simultaneously in focus at the objective sample plane. The Raman signal generated there was then collected by the objective and separated from the Rayleigh input beam using a fluorescence cube with a long-pass (Stokes Raman) Semrock RazorEdge dichroic beamsplitter chosen for the 514.52 nm $Ar^+$ laser line used in this study. The laser spot could then be focused on the sample region of interest while being viewed through the eyepieces or imaged at the camera port of the microscope using the viewport slider. A spatially positionable pinhole (~300 μm diameter) at the camera port was used to achieve confocality. The light cone exiting the pinhole expanded a focal length to lens f1 which collimated the beam for Rayleigh filtering with an angle-tuned Semrock LP02-514RE long pass filter. The beam was finally refocused by a lens f2 to the f/4 entrance aperture of a Chromex 250is spectrograph. With f1/f2 set to 5, the input image at the spectrograph of the 100-300 μm diameter camera port pinhole was minified 5x permitting most of the signal light to enter the 30-50 μm wide spectrograph entrance slit. Spectra were collected with a cooled SBIG ST-10 CCD camera located at the spectrograph image plane. The CCD camera with 2184 x 1470 pixels of area 6.9μm squared, provided 1-2 wavenumber resolution in the 2184 x-direction. Because of the small spot size at the spectrograph entrance slit, only 20-30 y-direction camera rows were co-added during binning of the signal to minimize dark count.

### 2.3 Light Treatment Methods

#### 2.3.1 UV treatment at 300 nm

A 100 watt PTI mercury arc lamp in combination with a Jarrell-Ash ¼ meter monochromator fitted with a ruled grating blazed for 250 nm, provided wavelength selection for sample irradiation at 300 nm. Entrance and exit slits for the monochromator were opened to several mms to provide adequate output intensity over the range 300±2 nm. After leaving the monochromator, the light beam was refocused to the sample location. The intensity there was determined using a calibrated Laser Precision Corp. radiometer, consisting of a RL-3610 Power Meter and a RKP-360 pyroelectric detector. Radiation dosage in mJ/cm$^2$ was determined by the product of intensity and exposure time. Once an *e. coli* cell suspension was freshly prepared, 100 µl of the cell suspension was pipetted into a cylindrical sterile glass cup for UV treatment. This cup had an 8 mm internal diameter and depth of 3 mm. After treatment, the cell samples were serial diluted in sterile phosphate buffer saline (PBS) and drop-plated on TSA plates for validation of the complete kill of the bacterial cells.

*2.3.2 Visible light treatment at 405nm*

Visible radiation at 405 nm was obtained from a 100 mW GaN laser diode beam expanded to illuminate the same concentration of cells in the cylindrical cup described above. The radiation intensity was determined using a Nova II Ophir laser power meter. Exposure at a fluence of 300 mW/cm$^2$ was continued until $2 \times 10^5$ mJ/cm$^2$, that is twice the lethal dose for this wavelength, was delivered to the sample. No heating of the sample occurred at this power density and the treated cells would not culture after treatment.

*2.4 Sample deposition and Spectra Collection*

Microscope ready samples were prepared by pipetting 0.3µl of cell suspension onto a clean CaF$_2$ microscope slide which was then placed on the microscope stage. Liquid evaporation was complete in less than 10 minutes and transmission illumination of the stage showed cells were distributed over a few square millimeters of plate area. Salt crystallites grew interspersed with the bacteria on the slide and were easily isolated from the laser measurement beam. Imaging with air immersion objectives of 60x or 100x revealed bacterial plate coverage to be approximately 20%. By manipulating the positions of cells under the focused laser input beam, one could place the beam spot on a group of 2-5 individual bacterial cells for spectral examination. Aside from increasing the signal several fold, placing several cells under the beam spot averaged possible variation in the signal response due to cell cycle variation in the culture harvest. Spectra were obtained using an input laser power of 2-5 mW for a collection period of up to 5 minutes. Dark background CCD counts were averaged over 60 exposures to improve the signal-to-noise. This background spectrum was subtracted from the spectral exposures. The CCD temperature was held at -10C for all accumulations. For a given sample, spectra from 3-4 different cell clusters were then averaged to improve signal-to-noise and further reduce any effects due to cell-cycle variation. This procedure produced reliable and reproducible results for all samples investigated.

## 3. Spectral Collection

*3.1 Reference spectra for E.coli in various stages of viability*

In order to catalog Raman spectra for comparison with those of the light treated samples, a series of JM109 *E.coli* with widely different conformations were studied. Fig. 2 (a) shows the Raman spectra for a sample of JM109 *e.coli* extracted from the culture in stationary phase growth, (b) shows that of a lyophilized version of the same sample, (c) that of a non-viable sample after storage at 0C in a refrigerator for one month and (d) a reference Raman spectrum of lyophilized *E.coli* DNA purchased from Sigma Aldrich Inc. Since spectra (a) and (b) are virtually identical, the evaporation drying process does not appear to modify the samples. The spectral peaks arising from DNA components are labeled D, a uracil peak near 815 cm$^{-1}$ U [10,21], amino acid/protein P, lipid L, and carbohydrates C. Fig. 2(c), the non-viable one month old sample contains little or no DNA/RNA signature and a halving of signal strength in the broad peaks centered at 1250 cm$^{-1}$ and 1330 cm$^{-1}$. This is of interest as these two peak groups are normally assumed to be due to proteins. Clearly, much of the signal in these regions for viable cells is due to DNA/RNA. The DNA spectrum of Fig. 2(d) does not include the ~815 cm$^{-1}$ peak due to uracil in the RNA component of all samples. The DNA Raman spectrum includes components from the phosphate backbone as well as that from the nucleic acids so individual peaks in the spectrum were not identified specifically. Table 2 summarizes Raman peak locations associated with molecular component structure(s).

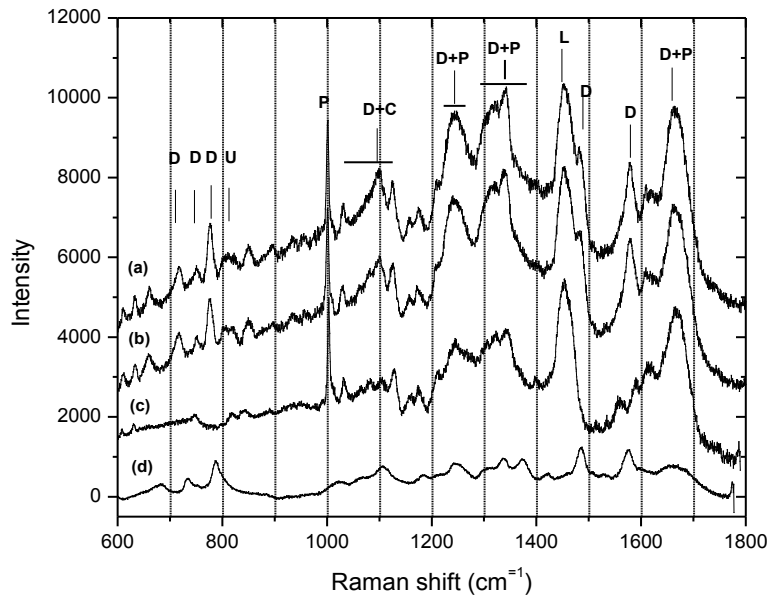

**Figure 2.** Confocal micro-Raman spectra of (a) freshly cultured stationary phase JM109 *E.coli,* (b) lyophilized version of the same batch, and (c) non-viable *E.coli* resulting from being stored at 0°C for 1 month, (d) lyophilized *E.coli* DNA strands purchased from Sigma Aldrich Inc. In spectrum (a) peaks labeled D are due to DNA, U due to uracil in RNA, P due to proteins, and C due to carbohydrate.

Table 1. Approximate macromolecular components [20] by percent dry weight and associated Raman spectral regions for the JM109 *E.coli* bacterium.

| Structural component | % dry weight | cm⁻¹ | cm⁻¹ | cm⁻¹ | cm⁻¹ |
|---|---|---|---|---|---|
| **DNA/RNA** | 23.6 | 670-820 | 1000-1125 | 1230-1280 | 1480,1570 |
| | | | | 1300-1360 | 1650-1700 |
| **Protein/amino** | 55 | 1004 Phe | 1220-1340 | 1440 | 1660-1710 |
| **Lipid** | 9.1 | 1440-1480 | | | |
| **Carbohydrate** | 8.4 | 1050-1150 | | | |

### 3.2 *Spectra for Bactericidal Treatment at 300 nm*

The *E.coli* cells were harvested after 14 hours of culturing and are therefore from the late exponential growth phase. During the 2x lethal treatment of 200 mJ/cm², the exposure level was 250 μW/cm² so no heating of the sample could occur during the several minute of exposure. Fig. 3 (a) and (b) show Raman spectra with dark count subtracted for untreated and 2x lethal dose exposure to 300 nm light. The spectra contained no fluorescence component. Curve (c) is the difference, and (d) is again the reference Raman spectrum of *E.coli* DNA.

Note that the difference spectrum is almost a duplicate of the DNA reference with the additional uracil peak near 815 cm$^{-1}$. This shows that the nucleic acid components of the spectrum are substantially reduced by these 4.13 eV photons since they are capable of bond cleaving throughout the cell system (see Table 1). Since the lethal fluence at this wavelength is so low (~100 mJ/cm$^2$), perhaps near resonance with the nucleic acid bond energies C-H (4.25 eV), N-H (4.03 eV), and P-O (3.90 eV), is responsible for the disproportionate reduction of the DNA/RNA Raman signal. It is also worth noting that a micro-Raman measurement using a sub-350 nm laser source of a few milliwatts and a focus spot of about 1 micron diameter, would exceed the lethal fluence at the objective focus by thousands of times in only seconds of measurement.

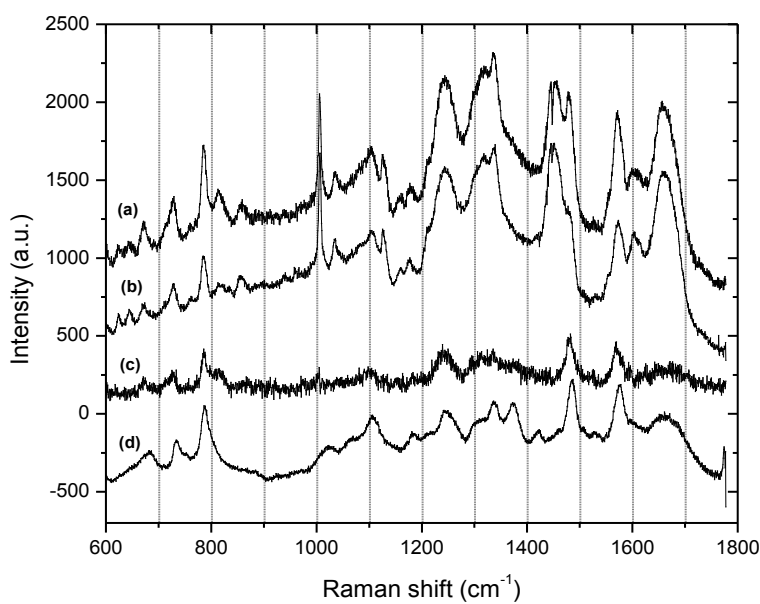

**Figure 3.** Confocal micro-Raman spectra of (a) as-grown live JM109 *E.coli*, (b) the same sample exposed to a 2x lethal fluence of 300 nm radiation, (c) the difference spectrum (a) - (b), and (d) a reference spectrum of *E.coli* DNA strands.

*3.3 Spectra for Treatment at 405 nm*

Fig. 4 (a) and (b) show Raman spectra for untreated and 2x lethal dose exposure to 405 nm light. Curve (c) is the difference, and (d) is again the reference Raman spectrum of *E.coli* DNA. The difference spectrum in this case shows no reduction in intensity in the DNA/RNA only regions but a significant reduction in the lipid peak intensity at 1440 cm$^{-1}$ and in the protein and protein plus DNA signal regions 1230-1265 cm$^{-1}$ and 1670 cm$^{-1}$. Since these 3.06 eV photons are close in energy to that of the C-N (3.12 eV) amino acid linking protein bonds, perhaps the protein change is reasonable. Why the lipid intensity decrease is so large is unclear as the C-C (3.60 eV) bond energy is significantly stronger than the incident light. Perhaps resonant C-P (3.04 eV) bonds in the membrane are responsible for the deterioration of the lipid structure.

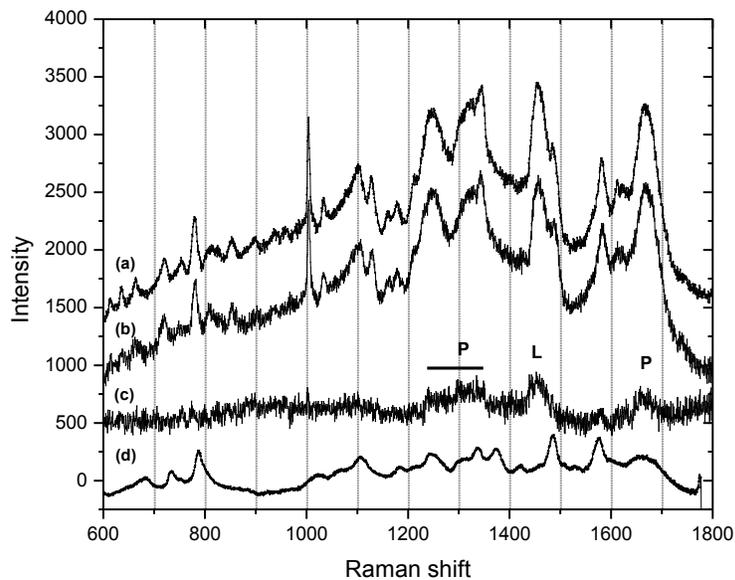

**Figure 4**. Confocal micro-Raman spectra of (a) as-grown live JM109 *E.coli*, (b) the *E.coli* exposed to a 2x lethal fluence of 405 nm light, (c) the difference spectrum (a) - (b) with protein and lipid peaks marked, and (d) the reference spectrum of *E.coli* DNA.

## 4. Discussion and Conclusions

Significant and quite different alterations of the Raman spectrum of live *E.coli* result after exposure to lethal fluences of 300 nm (UV-B) and 405 nm (violet) light. The energy per 300 nm photon is sufficient to damage most of the inter-atomic bonds in the bacterium and major change in the Raman spectrum, particularly in the RNA/DNA spectral regions is observed. By contrast, the 405 nm violet photon energy is insufficient to sever the stronger system bonds and the much larger lethal fluence required in this case produces little modification of the DNA/RNA region but significant intensity reduction in the protein and lipid spectral regions. We speculate that the changes for 300 nm exposure are due to near energy resonance of the light with C-P, N-H, P-O nucleotide and for 405 nm light with C-N protein plus C-P lipid phosphate membrane structures. The results provide a reminder that optical studies such as Raman or fluorescence imaging of live bacteria with sub-400nm light should be carried out with due consideration to lethal fluence issues. While both wavelengths in this study were lethal to the bacterium, the longer wavelength treatment should be less likely to induce mutation effects in live target species.

## Acknowledgements

This research was supported by Natural Science and Engineering Research Council of Canada Discovery Grant 6710-2006 to W.J.Keeler and Discovery Grant 238378-2010 to KT Leung.